\documentclass[amssymb,aps]{revtex4} 
\input{epsf}
\usepackage{graphicx}

\begin{document}

\title{A Relativistic Mean Field Model for Entrainment in General 
Relativistic Superfluid Neutron Stars}

\author{G. L. Comer}

\affiliation{Department of Physics, Saint Louis University, 
St.~Louis, MO, 63156-0907, USA}

\author{R. Joynt}

\affiliation{Department of Physics, University of Wisconsin - 
Madison, Madison, WI 53706, USA}

\date{\today}
	

\def\beq{\begin{equation}}
\def\eeq{\end{equation}}
\def\n{n}
\def\p{p}
\def\d{\delta}
\def\A{{\cal A}}
\def\B{{\cal B}}
\def\C{{\cal C}}
\def\a00{{{\cal A}_0^0}}
\def\b00{{{\cal B}_0^0}}
\def\c00{{{\cal C}_0^0}}
\def\D00{{{\cal D}_0^0}}
\def\d{\delta}
\def\A{{\cal A}}
\def\B{{\cal B}}
\def\C{{\cal C}}
\def\M{{\cal M}}
\def\wh{\widehat}
\def\wa{\widetilde}
\def\ha{\overline}
\def\gon{{\gamma^\n_0}}
\def\gop{{\gamma^\p_0}}
\def\gzn{{\gamma^z_\n}}
\def\gzp{{\gamma^z_\p}}

\begin{abstract}
General relativistic superfluid neutron stars have a significantly 
more intricate dynamics than their ordinary fluid counterparts. \ 
Superfluidity allows different superfluid (and superconducting) 
species of particles to have independent fluid flows, a consequence 
of which is that the fluid equations of motion contain as many fluid 
element velocities as superfluid species. \ Whenever the particles 
of one superfluid interact with those of another, the momentum of 
each superfluid will be a linear combination of both superfluid 
velocities. \ This leads to the so-called entrainment effect whereby 
the motion of one superfluid will induce a momentum in the other 
superfluid. \ We have constructed a fully relativistic model for 
entrainment between superfluid neutrons and superconducting protons 
using a relativistic $\sigma - \omega$ mean field model for the 
nucleons and their interactions. \ In this context there are two 
notions of ``relativistic'': relativistic motion of the individual 
nucleons with respect to a local region of the star (i.e.~a fluid 
element containing, say, an Avogadro's number of particles), and the 
motion of fluid elements with respect to the rest of the star. \ 
While it is the case that the fluid elements will typically maintain 
average speeds at a fraction of that of light, the supranuclear 
densities in the core of a neutron star can make the nucleons 
themselves have quite high average speeds within each fluid element. 
\ The formalism is applied to the problem of slowly-rotating 
superfluid neutron star configurations, a distinguishing 
characteristic being that the neutrons can rotate at a rate different 
from that of the protons.
\end{abstract}

\maketitle

\section{Introduction}

A new generation of gravitational wave detectors (LIGO, VIRGO, etc) 
are now working to detect gravitational waves from compact objects, 
such as black holes and neutron stars. \ With this detection we 
expect to have a unique probe of the physics that dictates their 
behavior. \ This is ushering in a new era where strong-field 
relativistic effects will play an increasingly important role. \ Only 
through their inclusion can we hope to accurately decipher what 
gravitational wave data will have to tell us. \ With that in mind, 
we present here a fully relativistic model of the so-called 
entrainment effect (to be described in some detail below) that is a 
necessary feature of the dynamics of superfluid neutron stars. 

For the densities appropriate to neutron stars there are attractive 
components of the strong force that should lead, via BCS-like 
mechanisms, to nucleon superfluidity and superconductivity. \ Indeed, 
calculations of supra-nulcear gap energies consistently lead to the 
conclusion that superfluid neutrons should form in the inner crust of 
a mature neutron star, with superfluid neutrons and superconducting 
protons in the core. \ Even more exotic possibilities have been 
suggested, such as pion condensates, superfluid hyperons, and 
superconducting quark matter. \ Perhaps most important is the 
well-established glitch phenomenon in pulsars the best description of 
which is based on superfluidity and quantized vortices. \ 
Superfluidity should affect gravitational waves from neutron stars 
by modifying the rotational equilibria and the modes of 
oscillations that these objects support \cite{AC01c,AC01b,GC02}.  

The success of superfluidity in describing the glitch phenomena is 
due in part to the fact that the superfluid neutrons of the inner 
crust represent a component that can move freely (for certain 
timescales) from the rest of the star. \ Explaining the glitch 
phenomena then becomes a question of how to transfer angular momentum 
between the various ``rotationally decoupled'' components. \ For the 
modes of oscillation, it is by now well established that a similar 
``decoupling,'' this time between the superfluid neutrons of the 
inner crust and core and a conglomerate of the remaining charged 
constituents (e.g.~crust nuclei, core superconducting protons, and 
crust and core electrons), leads to a mode spectrum for superfluid 
neutron stars that is quite different from that of their ordinary 
fluid counterparts (see \cite{GC02}, and references therein, for a 
complete review).  

Several recent studies \cite{AC01a,PCA02,ACL02,ACP02a,ACP02b} have 
established that the entrainment effect is an important element in 
modelling the rotational equilibria and modes of oscillation of 
superfluid neutron stars. \ Sauls \cite{S89} describes the 
entrainment effect as a result of the quasiparticle nature of the 
excitation spectrum of the superfluid and superconducting nucleons. \ 
That is, the bare neutrons (or protons) are accompanied by a 
polarization cloud containing both neutrons and protons. \ Since both 
types of nucleon contribute to the cloud the momentum of the neutrons 
is modified so that it is a linear combination of both the neutron 
and proton particle number density currents, and similarly for the 
proton momentum. \ Thus when one species of nucleon acquires 
momentum, both types of nucleons will begin to flow.

In the core of a neutron star, the Fermi energies of nucleons (as 
well as some of the leptons) can become comparable to their 
mass-energies, because the Fermi energies are a function of the local 
particle number densities, and these can be quite high. \ This 
implies that any Newtonian model for entrainment must become less 
reliable as one probes deeper into the core of a neutron star, and 
thus a relativistic formulation is required. \ In fact, we will see 
that the Newtonian parameterized model of Prix et al.~\cite{PCA02} 
does deviate most from the relativistic model in the core. \ There 
are two purposes for which a relativistic formulation is necessary. 
\ At the microscopic level, the nucleons will (locally) have average 
speeds that are comparable to the speed of light. \ As well, at a 
mesoscopic level, the fluid elements, which contain a large number of 
nucleons, could have average speeds that are also comparable to the 
speed of light. \ The formalism that we develop here will be 
relativistic in both respects. \ One should note, though, that in 
realistic astrophysical scenarios (e.g.~when an isolated neutron star 
undergoes linearized oscillations, or a pulsar exhibits a glitch) the 
fluid element average speeds are typically only a few percent of that 
of light.  

To date, studies of superfluid dynamics in neutron stars have 
relied on models of entrainment that are obtained in the Newtonian 
regime. \ For instance, a few of the most recent studies 
\cite{LM00,ACL02} have employed a parameterized model for 
entrainment that is inspired by the Newtonian, Fermi-liquid 
calculations of Borumand et al.~\cite{BJK96}. \ An alternative 
formulation \cite{PCA02}---motivated by mathematical simplicity that 
allows for analytic solutions for slowly rotating Newtonian 
superfluid neutron stars---for parameterizing entrainment has been 
recently put forward. \ Here we take a different approach, and this 
is to use a $\sigma - \omega$ relativistic mean field model, of the 
type that is described in detail by Glendenning \cite{G97}. \ 
Although a relativistic Fermi-liquid formalism exists \cite{CB76}, we 
prefer to use the mean field model because it is sufficiently simple 
that semi-analytical formulas result, and a clear connection between 
the coupling parameters at the microscopic level can be made to the 
macroscopic properties (such as mass and radius) of the star. \ An 
immediate consequence is the ability to compare the relativistic 
entrainment model with the two parameterized models. \ We will see 
that the model used by Prix et al.~\cite{PCA02}, although limited, is 
a better fit than the other formulation.

The next section begins with a review of the $\sigma - \omega$ model. 
\ That is followed by an application of the mean field approximation 
to obtain an equation of state that includes entrainment. \ In 
Sec.~3, we briefly review the general relativistic superfluid 
formalism and how it is used to describe slowly rotating 
configurations. \ We then use the mean field results to produce 
explicit models. \ After some concluding remarks, an appendix is 
given that contains some of the technical details and results. \ 
Throughout we will use ``MTW'' \cite{MTW} conventions, a consequence 
of which is that several equations will have minus sign differences 
with, for instance, those of \cite{G97}.

\section{Relativistic Mean Field Theory of Coupled Fluids}

To create a seamless conceptual basis for general relativistic 
calculations of dynamic processes in neutron stars, we need a 
covariant formalism that describes the strongly interacting coupled 
neutron and proton fluids. \ It should be sufficiently simple that 
it provides physical insight, yet accurate enough that it can serve 
as the basis for realistic numerical calculations. \ For static 
stars, this role is played by the $\sigma - \omega$ effective 
mean-field theory \cite{G97}. \ Our task in this paper is to 
generalize this theory to dynamic stars. \ In particular, we are 
interested in situations where there is relative motion of the two 
fluids, since the entrainment of one by the other turns out to play 
a large role in the dynamics.

The Lagrangian density for the baryons and the mesons that the baryons
exchange is as in the static case. \ It is \ 
\begin{equation}
    L = L_{b} + L_{\sigma } + L_{\omega } + L_{int} \ ,
\end{equation}%
with%
\begin{equation}
   L_{b} = \bar{\psi} (i\gamma _{\mu } \partial^{\mu } - m) \psi 
\end{equation}%
as the baryon Lagrangian. \ Here $\psi $ is an 8-component spinor 
with the proton components as the top 4 and the neutron components 
as the bottom 4. \ The $\gamma _{\mu }$ are the corresponding 
$8 \times 8$ block diagonal Dirac matrices. \ The Lagrangian for the  
$\sigma$ mesons is 
\begin{equation}
    L_{\sigma } = - \frac{1}{2}\partial _{\mu } \sigma 
                  \partial^{\mu }\sigma -\frac{1}{2}m_{\sigma }^{2} 
                  \sigma ^{2} \ . 
\end{equation}%
The Lagrangian for the $\omega$ mesons is 
\begin{equation}
    L_{\omega } = - \frac{1}{4}\omega _{\mu \nu } \omega ^{\mu \nu } 
             - \frac{1}{2} m_{\omega }^{2}\omega _{\mu }\omega ^{\mu }
\end{equation}%
where $\omega _{\mu \nu } = \partial _{\mu }\omega _{\nu } - 
\partial _{\nu}\omega _{\mu }.$ \ The interaction Lagrangian density 
is%
\begin{equation}
    L_{int} = g_{\sigma } \sigma \bar{\psi} \psi - g_{\omega } 
               \omega _{\mu }\bar{\psi} \gamma^{\mu }\psi \ .
\end{equation}%
The Euler-Lagrange equations are%
\begin{eqnarray}
    \left(- \Box + m_{\sigma }^{2}\right) \sigma &=& g_{\sigma } 
    \bar{\psi}\psi \ , \\ 
    && \cr
    \left(- \Box + m_{\omega }^{2}\right) \omega _{\mu } + 
    \partial _{\mu }\partial^{\nu } \omega _{\nu } &=& - g_{\omega } 
    \bar{\psi} \gamma _{\mu } \psi \ , \\ 
    && \cr
   (i \gamma _{\mu } \partial^{\mu } - m) \psi &=& g_{\omega } 
   \gamma _{\mu }\omega^{\mu }\psi -g_{\sigma }\sigma \psi \ .
\end{eqnarray}%
Finally, the stress-energy tensor takes the form
\beq 
    T^{\mu \nu } = T_{b}^{\mu \nu } + T_{\sigma}^{\mu \nu } + 
    T_{\omega }^{\mu \nu }+T_{int}^{\mu \nu }
\eeq
containing contributions from the baryons $(b)$, the mesons 
$(\sigma ,\omega )$, and the interaction. \ Individually, these are
\begin{eqnarray}
    T_{b}^{\mu \nu } &=& - i \bar{\psi} (\gamma^{\mu} 
    \partial^{\nu} - \eta^{\mu \nu} \gamma^{\alpha} 
    \partial_{\alpha}) \psi - m \eta^{\mu \nu} \bar{\psi} \psi \ , \\
    && \cr
    T_{\sigma}^{\mu \nu} &=& \partial^{\mu} \sigma \partial^{\nu} 
    \sigma - \frac{1}{2} \eta^{\mu \nu} 
    m_{\sigma}^{2}\sigma^{2} - \frac{1}{2} \eta^{\mu \nu} 
    \partial^{\alpha} \sigma \partial_{\alpha} 
    \sigma \ , \\
    && \cr
    T_{\omega}^{\mu \nu} &=& \left(\partial^{\mu} \omega^{\alpha} - 
    \partial^{\alpha} \omega^{\mu} \right) \partial^{\nu} 
    \omega_{\alpha} - \frac{1}{2} \eta^{\mu \nu} m_{\omega}^{2} 
    \omega^{\alpha} \omega _{\alpha} - \frac{1}{4} \eta^{\mu \nu} 
    m_{\omega}^{2} \omega^{\alpha \beta} \omega _{\alpha \beta} \ , \\
    && \cr
    T_{\iota nt}^{\mu \nu} &=& \eta^{\mu \nu} g_{\sigma} \sigma 
    \bar{\psi}\psi - \eta ^{\mu \nu} g_{\omega} \omega _{\alpha} 
    \bar{\psi}\gamma^{\alpha} \psi \ .
\end{eqnarray}

We now solve these equations in the mean field approximation, 
eventually in a frame in which the neutrons have zero spatial 
momentum while the protons have on average a wavevector 
$K_{\mu} = (K_{0},0,0,K_{z}).$ \ In this approximation we ignore all 
gradients of the averaged sigma and omega fields, and the neutrons 
and protons are taken to be in plane-wave states. \ The problem 
simplifies considerably and we find for the $\sigma$ and 
$\omega_{\mu}$ fields and the stress-energy tensor $T^{\mu}_{\nu}$ 
that
\begin{eqnarray}
    m_* &=& m - c_{\sigma}^{2} \left\langle \bar{\psi} \psi 
    \right\rangle \ , \\
    && \cr
    \left\langle g_{\omega} \omega _{\mu} \right\rangle  &=& 
    - c_{\omega}^{2} \left\langle \bar{\psi} \gamma _{\mu} \psi 
    \right\rangle \ , \\ 
    && \cr
    \left\langle T^{\mu}_{\nu} \right\rangle &=& - \frac{1}{2} 
    \left(c_{\omega}^{-2} \left\langle g_{\omega} \omega^{\alpha} 
    \right\rangle \left\langle g_{\omega} \omega_{\alpha} 
    \right\rangle + c_{\sigma}^{- 2} \left[m - m_*\right]^2\right) 
    \delta^{\mu}_{\nu} - i \left\langle \bar \Psi \gamma^{\mu} 
    \partial_{\nu} \Psi \right\rangle \ ,
\end{eqnarray}
where, for later convenience, we have introduced the notation 
$c^2_{\sigma} = (g_{\sigma}/m_{\sigma})^2$ and $c^2_{\omega} = 
(g_{\omega}/m_{\omega})^2$ and the Dirac effective mass $m_*$, i.e.
\beq
    \left\langle g_{\sigma} \sigma \right\rangle = m - m_*\ .
\eeq

Restricting to the zero-momentum frame of the neutrons leads to a 
set of algebraic equations for the $\omega_{\mu}$ field:%
\begin{eqnarray}
    \left\langle g_{\omega} \omega _{0} \right\rangle &=& 
    c_{\omega}^{2} \left\langle \bar{\psi} \gamma _{0} \psi 
    \right\rangle \ , \\
    && \cr
    \left\langle g_{\omega} \omega _{z} \right\rangle &=& 
    c_{\omega}^{2} \left\langle \bar{\psi} \gamma _{z} \psi 
    \right\rangle \ .
\end{eqnarray}%
The final equation is not needed in the case where both neutrons and 
protons have zero average momentum, since $\left\langle \omega _{z} 
\right\rangle$ then vanishes by isotropy. \ In this case, the 
neutrons and protons have a common rest frame and  $\left\langle 
\bar{\psi} \gamma^{0} \psi \right\rangle = \psi^{\dagger} \psi = \n + 
\p$ where $\n$ and $\p$ are the baryon number densities of the 
neutrons and protons, respectively. \ The addition of the spatial 
velocity component complicates the solution of the problem 
considerably, in part because there is no longer a common rest frame 
for all the baryons. \ Each expectation value on the RHS of these 
equations involves an integration over the Fermi spheres of the 
particles, whose radii can be shown (c.f.~the next section) to be 
$k_{\n}=\left(3 \pi^{2} \n^0\right)^{1/3}$ and $k_{\p}=\left(3 
\pi^{2} \p^0\right)^{1/3}$, where $\n^0$ ($\p^0$) is the 
zero-component of the conserved neutron (proton) number density 
current $\n^{\mu}$ ($\p^{\mu}$). \ The proton Fermi surface is 
displaced by $K_{z}\widehat{z}$. \ We are interested in the case 
$K_{z} << k_{\n},k_{\p}$, but the expressions for general $K_{z}$ are 
not more complicated than the power series expansion.

Noting that 
\begin{equation}
   \left\langle \bar{\psi} \left[\gamma^{\mu} (i \partial_{\mu} -
    g_{\omega} \omega_{\mu}) - m_*\right] \psi \right\rangle = 0
\end{equation}%
we find%
\begin{equation}
    \left(k^{0} + g_{\omega} \omega^{0}\right)^{2} = \left(\vec{k} + 
    g_{\omega} \omega^{z} \widehat{z}\right)^{2} + m_*^{2} \ , 
\end{equation}%
where we have dropped expectation value brackets for the mean 
values of the fields. \ The energy $\varepsilon $ of a baryon in a 
plane-wave state is given by%
\begin{equation}
     \varepsilon(\vec{k}) = E(\vec{k}) - g_{\omega} \omega^{0} = 
     \sqrt{\left(\vec{k} + g_{\omega} \omega^{z} \widehat{z}
     \right)^{2} + m_*^{2}} - g_{\omega}\omega^{0} \ .
\end{equation}%
Thus we see that $\omega _{0}$ contributes a constant shift, 
$\omega _{z}$ gives a preferred frame for the momenta, and $\sigma$ 
renormalizes the mass to the Dirac mass.

As an example of how the expectation values are evaluated, we give the
scalar density (letting $K_z = K$, for ease of notation):%
\begin{eqnarray}
    \left\langle \bar{\psi}\psi \right\rangle  &=& \frac{1}{(2 
    \pi)^{3}} \int_{occ} d^{3}k \,\frac{\partial E}{\partial m} \\
    && \cr
    &=& \frac{2}{\left(2 \pi\right)^{3}} \int_{|\vec{k}|<k_{\n}}\,
    d^{3}k\,\frac{m_*}{\sqrt{(\vec{k} + g_{\omega} \omega _{z} 
    \widehat{z})^{2} + m_*^{2}}} + \frac{2}{\left(2 \pi\right)^{3}} 
    \int_{|\vec{k} - K \widehat{z}| < k_{\p}}\,d^{3}k\,
    \frac{m_*}{\sqrt{(\vec{k} + g_{\omega} \omega_{z} 
    \widehat{z})^{2} + m_*^{2}}} \\
    && \cr
    &=&\frac{2}{\left(2 \pi\right)^{3}} \int_{|\vec{k}| < k_{\n}}
    \,d^{3}k\,\frac{m_*}{\sqrt{(\vec{k} + g_{\omega} \omega_{z} 
    \widehat{z})^{2} + m_*^{2}}} + \frac{2}{\left(2 \pi\right)^{3}} 
    \int_{|\vec{k}| < k_{p}} \,d^{3}k\,\frac{m_*}{\sqrt{(\vec{k} + 
    g_{\omega} \omega _{z} \widehat{z} + K \widehat{z})^{2} + 
    m_*^{2}}} \ ,
\end{eqnarray}%
and the average four-velocity components of the baryons: 
\begin{eqnarray}
    \left\langle \bar{\psi} \gamma^{0} \psi \right\rangle &=& 
    \frac{1}{(2 \pi)^{3}} \int_{occ}\,d^{3}k\,
    \frac{\partial E}{\partial k_0} \cr
    && \cr
    &=&\frac{2}{\left(2 \pi \right)^{3}} \int_{|\vec{k}| < k_{\n}}
    \,d^{3}k\, + \frac{2}{\left(2 \pi\right)^{3}} \int_{|\vec{k} - K 
    \widehat{z}| < k_{\p}} \,d^{3}k\, \cr
    && \cr
    &=&\frac{2}{\left(2 \pi\right)^{3}} \int_{|\vec{k}| < k_{\n}}
    \,d^{3}k\, + \frac{2}{\left(2 \pi\right)^{3}} \int_{|\vec{k}| < 
    k_{\p}} \,d^{3}k \ , \\
    && \cr
    \left\langle \bar{\psi} \gamma^{z} \psi \right\rangle &=& 
    \frac{1}{(2 \pi)^{3}} \int_{occ}\,d^{3}k\,
    \frac{\partial E}{\partial k_{z}} \cr
    && \cr
    &=&\frac{2}{\left(2 \pi \right)^{3}} \int_{|\vec{k}| < k_{\n}}
    \,d^{3}k\,\frac{k^{z} + g_{\omega} \omega^{z}}{\sqrt{(\vec{k} + 
    g_{\omega} \omega^{z} \widehat{z})^{2} + m_*^{2}}} + 
    \frac{2}{\left(2 \pi\right)^{3}} \int_{|\vec{k} - K \widehat{z}| 
    < k_{\p}} \,d^{3}k\,\frac{k^{z} + g_{\omega} \omega^{z}}{\sqrt{(
    \vec{k} + g_{\omega} \omega^{z} \widehat{z})^{2} + m_*^{2}}} \cr
    && \cr
    &=&\frac{2}{\left(2 \pi\right)^{3}} \int_{|\vec{k}| < k_{\n}}
    \,d^{3}k\frac{k^{z} + g_{\omega} \omega^{z}}{\sqrt{(\vec{k} + 
    g_{\omega} \omega^{z} \widehat{z})^{2} + m_*)^{2}}} + 
    \,\frac{2}{\left(2 \pi\right)^{3}} \int_{|\vec{k}| < k_{\p}} 
    \,d^{3}k\frac{k^{z} + g_{\omega} \omega^{z} + K}{\sqrt{(\vec{k} 
    + g_{\omega} \omega^{z} \widehat{z} + K \widehat{z})^{2} + 
    m_*^{2}}} \ .
\end{eqnarray}%
Thus we have reduced the problem to a set of nonlinear equations for 
the $m_* $, $\omega _{0}$, and $\omega _{z}$ fields that must be 
solved numerically. \ This can be done for any set of the input 
parameters $k_{\n}$, $k_{\p}$, and $K$. \ The interaction and mass 
parameters for the effective fields have been determined from nuclear 
physics, and they are discussed further below. \ Once this is done, 
we still need expressions for the stress-energy tensor, which is the 
input for the Einstein equations.

Again, specializing to the zero-momentum frame of the neutrons, the 
only non-zero stress-energy tensor components are%
\begin{eqnarray}
    \left\langle T^{0}_{0} \right\rangle &=& - {1 \over 2} 
    c_{\omega}^2 \left(\left\langle \bar{\psi} \gamma^{0} \psi 
    \right\rangle^2 - \left\langle \bar{\psi} \gamma^{z} \psi 
    \right\rangle^2\right) - {1 \over 2} c_{\sigma}^{- 2} \left(m^2 
    - m_*^2\right) - \left\langle \bar{\psi} \gamma^i k_i \psi 
    \right\rangle \ , \\
    && \cr
    \left\langle T^{0}_{z} \right\rangle &=& \left\langle \bar{\psi} 
    \gamma^{0} k_{z} \psi \right\rangle \ , \\
    && \cr
    \left\langle T^{x}_{x} \right\rangle &=& \left\langle T^{y}_{y} 
    \right\rangle = {1 \over 2} c_{\omega}^2 \left(\left\langle 
    \bar{\psi} \gamma^{0} \psi \right\rangle^2 - \left\langle 
    \bar{\psi} \gamma^{z} \psi \right\rangle^2\right) - {1 \over 2} 
    c_{\sigma}^{- 2} \left(m - m_*\right)^2 + \left\langle \bar{\psi} 
    \gamma^{x} k_{x} \psi \right\rangle \ , \\
    && \cr
    \left\langle T^{z}_{z} \right\rangle &=& {1 \over 2} 
    c_{\omega}^2 \left(\left\langle \bar{\psi} \gamma^{0} \psi 
    \right\rangle^2 - \left\langle \bar{\psi} \gamma^{z} \psi 
    \right\rangle^2\right) - {1 \over 2} c_{\sigma}^{- 2} \left(m - 
    m_*\right)^2 + \left\langle \bar{\psi} \gamma^{z} k_{z} \psi 
    \right\rangle \ . 
\end{eqnarray}%
Some of the expressions have been simplified using the equations of 
motion. 

Each component of $\left\langle T^{\mu}_{\nu} \right\rangle$ again 
involves an integration over the Fermi surfaces, but now in terms of 
completely known parameters. \ For example, to determine 
$\left\langle T^{z}_{z} \right\rangle$, we need%
\begin{eqnarray}
    \left\langle \bar{\psi} \gamma^{z} k_{z} \psi \right\rangle &=& 
    \frac{2}{(2 \pi)^{3}} \int_{|\vec{k}| < k_{\n}}d^{3}k\,\,k^{z} 
    \frac{\partial E}{\partial k^{z}} + \frac{2}{(2 \pi)^{3}} 
    \int_{|\vec{k} - K \widehat{z}| < k_{\p}}d^{3}k\,k^{z} 
    \frac{\partial E}{\partial k^{z}} \cr
    && \cr
    &=& \frac{2}{(2\pi )^{3}}\int_{|\vec{k}| < k_{\n}}d^{3}k\,\,
    k_{z} \left[k_{z} + g_{\omega} \omega_{z}\right] \left[
    \left(\vec{k} + g_{\omega} \omega^{z} \widehat{z}\right)^{2} + 
    m_*^{2}\right]^{- 1/2} + \cr
    && \cr
    &&\frac{2}{(2\pi)^{3}} \int_{|\vec{k}| < k_{\p}}d^{3}k
    \,\,\left[k_{z} + K\right] \left[k_{z} + g_{\omega} \omega_{z} 
    + K\right] \,\left[\left(\vec{k} + g_{\omega} \omega^{z} 
    \widehat{z} + K \widehat{z}\right)^{2} + m_*^{2}\right]^{- 1/2} 
    \ ,
\end{eqnarray}
and for $\left\langle T^{0}_{z} \right\rangle$
\beq
    \left\langle \bar{\psi} \gamma^{0} k_{z} \psi \right\rangle = 
    \frac{2}{(2 \pi)^{3}} \int_{|\vec{k}| < k_{\n}}d^{3}k\,\,k_{z} 
    + \frac{2}{(2 \pi)^{3}} \int_{|\vec{k}| < k_{\p}}d^{3}k\,
    \left(k_{z} + K\right) = {k^3_\p \over 3 \pi^2} K \ .
\eeq

The main result of this section is thus a well-defined prescription 
for producing the functions 
$\left\langle T^{\mu}_{\nu}\right\rangle(k_{\n},k_{\p},K_{z})$. 
\ In the next section we take this prescription and produce from it 
the so-called master function, including entrainment, that is used in 
the general relativistic superfluid field equations.

\section{General Relativistic Superfluid Formalism}

The formalism to be used here, and motivation for it, has been 
described in great detail elsewhere 
\cite{C89,CL94,CL95,CL98a,CL98b,LSC98,CLL99,P00,AC01c,GC02}, and so 
we will review only the highlights. \ The central quantity of the 
superfluid formalism is the master function $\Lambda$. \ It 
depends on the three scalars $\n^2 = - \n_\mu \n^\mu$, $\p^2 = - 
\p_\mu \p^\mu$ and $x^2 = -p_\mu n^\mu$ that can be formed from the 
conserved neutron ($n^{\mu}$) and proton ($p^{\mu}$) number density 
currents. \ Furthermore, the master function is such that $- 
\Lambda(\n^2,\p^2,x^2)$ corresponds to the total thermodynamic energy 
density if the neutrons and protons flow together (as measured in the 
comoving frame). \ Once the master function is provided the 
stress-energy tensor is given by
\beq
    T^\mu_\nu = \Psi \delta^\mu_\nu + \n^\mu \mu_\nu + \p^\mu 
    \chi_\nu \ ,
\eeq
where
\beq
    \Psi = \Lambda - n^\rho \mu_\rho - p^\rho \chi_\rho \label{press}
\eeq
is the generalized pressure, and
\begin{eqnarray}
    \mu_\nu &=& {\cal B} n_\nu + {\cal A} p_\nu  \ , \\
    \chi_\nu &=& {\cal A} n_\nu + {\cal C} p_\nu \ , 
\end{eqnarray}
are the chemical potential covectors. \ We also have
\beq
    {\cal A} = -{ \partial \Lambda \over \partial x^2} \ , \quad 
    {\cal B} = -2{ \partial \Lambda \over \partial n^2} \ , \quad 
    {\cal C} = -2 { \partial \Lambda \over \partial p^2} \ . \quad 
\eeq
The momentum covectors $\mu_\nu$ and $\chi_\nu$ are dynamically, and 
thermodynamically, conjugate to $n^\nu$ and $\p^\nu$ and their 
magnitudes are the chemical potentials of the neutrons and the 
protons, respectively. \ The two covectors also make manifest the 
entrainment effect; that is, we see that the momentum of one 
constituent ($\mu_\nu$, say) carries along some of the mass current 
of the other constituent ($\mu_\nu$ is a linear combination of 
$n_\nu$ and $p_\nu$). \ We can also see that there is no 
entrainment unless the master function depends on $x^2$. 

The field equations for this system take the form of two conservation 
equations for the neutrons and protons, i.e.
\beq
    \nabla_{\mu} \n^{\mu} = 0 \quad , \quad 
    \nabla_{\mu} \p^{\mu} = 0 \ ,
\eeq
which is a reasonable approximation given that the weak interaction 
timescale is much longer than the dynamical timescale of neutron 
stars for small amplitude deviations from equilibrium \cite{E88}, 
and two Euler equations, i.e.
\beq
    \n^{\mu} \nabla_{[\mu} \mu_{\nu]} = 0 \quad , \quad 
    \p^{\mu} \nabla_{[\mu} \chi_{\nu]} = 0 \ , 
\eeq
where the square braces means antisymmetrization of the enclosed 
indices.

\subsection{Extracting the Master Function from the Mean Field 
Results}

The two scales that enter this problem are the microscopic, on the 
scale of the nucleons, and the mesoscopic where one speaks in terms 
of the two interpenetrating superfluids. \ The fundamental 
``particles'' at the fluid level are the fluid elements which 
contain, say, an Avogadro's number worth of nucleons. \ The 
connection between the micro- and meso-scopic levels is via the 
averaged stress-energy components calculated earlier. \ Consider a 
fluid element deep in the core of the neutron star and orient the 
local coordinate frame in such a way that the z-axis of the frame is 
in the same direction as the proton momentum with respect to the 
neutrons. \ As shown just below, a unique combination of the averaged 
stress-energy components determined via the mean field theory will 
yield the master function. \ As this quantity is a scalar, the 
functional relationship we obtain between $\Lambda$ and the two 
particle number densities and the relative velocity of the protons 
with respect to the neutrons can then be applied anywhere in the star.

The key idea is to use the (local) relationship
\beq
    \left\langle T^{\mu}_{\nu} \right\rangle = \Psi \delta^\mu_\nu 
                 + \n^\mu \mu_\nu + \p^\mu \chi_\nu 
\eeq
to obtain $\Lambda$. \ In the perfect fluid case, the identification 
is made immediate by the fact that there is only one four-velocity 
$u^{\mu}$ for the system, and hence a preferred rest-frame for the 
particles. \ The local energy density of the fluid is thus uniquely 
obtained from $\Lambda = - \left\langle T^{\mu}_{\nu} \right\rangle 
u^{\nu} u_{\mu}$. \ In the superfluid case there are two 
four-velocities, and thus no preferred rest-frame. \ Fortunately, we 
can still obtain $\Lambda$ in a unique, and covariant, way, by using 
the trace $\left \langle T \right\rangle \equiv \left\langle 
T^{\mu}_{\mu} \right\rangle$ and the three scalars that can be formed 
from contracting $\left\langle T^{\mu}_{\nu} \right\rangle$ 
with $\n^{\mu}$ and $\p^{\mu}$, i.e.~$\left\langle T^{\mu}_{\nu} 
\right\rangle \n_{\mu} \n^{\nu}$, $\left\langle T^{\mu}_{\nu} 
\right\rangle \n_{\mu} \p^{\nu}$, and $\left\langle T^{\mu}_{\nu} 
\right\rangle \p_{\mu} \p^{\nu}$. \ We thus find that $\Lambda$ is 
given by
\beq
    \Lambda = - {1 \over 2} \left \langle T \right\rangle + 
              {3 \over 2} \left(x^4 - \n^2 \p^2\right)^{- 1} \left(
              \n^2 \p^2 \left[{1 \over \n^2} n^{\mu} \n^{\nu} + 
              {1 \over \p^2} \p^{\mu} \p^{\nu}\right] - x^2 \left[
              \n^{\mu} \p^{\nu} + \p^{\mu} \n^{\nu}\right]\right) 
              \left\langle T_{\mu \nu} \right\rangle \ ,
\eeq  
and the generalized pressure is
\beq
    \Psi = {1 \over 3} \left(\left\langle T \right\rangle - 
    \Lambda\right) \ .
\eeq
In like manner we find that 
\begin{eqnarray}
    \A &=& - \left(\n_{\mu} \p^{\nu} \left\langle 
           T^{\mu}_{\nu} \right\rangle + x^2 \Lambda\right) / 
           \left(x^4 - \n^2 \p^2\right) \ , \cr
       && \cr
    \B &=& \left(\p_{\mu} \p^{\nu} \left\langle 
           T^{\mu}_{\nu} \right\rangle + \p^2 \Lambda\right) / 
           \left(x^4 - \n^2 \p^2\right) \ , \cr
       && \cr
    \C &=& \left(\n_{\mu} \n^{\nu} \left\langle 
           T^{\mu}_{\nu} \right\rangle + \n^2 \Lambda\right) / 
           \left(x^4 - \n^2 \p^2\right) \ . \label{alg}
\end{eqnarray}

One other necessary component of uniting the mean field theory with 
the superfluid formalism is to relate (locally) $\n^{\mu}$ and 
$\p^{\mu}$ to the mean particle flux of the neutrons and protons; i.e.
\begin{eqnarray}
    \n^{\mu} &\equiv& \n u^{\mu}_{\n} = \left \langle 
                      \bar{\psi}_{\n} \gamma^{\mu} \psi_{\n} \right 
                      \rangle \ , \cr
             && \cr
    \p^{\mu} &\equiv& \p u^{\mu}_{\p} = \left \langle 
                      \bar{\psi}_{\p} \gamma^{\mu} \psi_{\p} \right 
                      \rangle \ , \label{coordefs}  
\end{eqnarray}
where $\psi_{\n}$ and $\psi_{\p}$ are the neutron and proton, 
respectively, components of the Dirac spinor $\psi$. \ Recall again 
that we have arranged that the average neutron and proton particle 
fluxes are in the z-direction. \ Thus, the unit vectors have only two 
components:
\begin{eqnarray}
    u^{\mu}_{\n} &=& \left\{u^0_{\n},0,0,u^3_{\n}\right\} \ , 
                     \ u^0_{\n} = \sqrt{1 + \left(u^3_{\n}\right)^2} 
                     \ , \cr
                 && \cr
    u^{\mu}_{\p} &=& \left\{u^0_{\p},0,0,u^3_{\p}\right\} \ , 
                     \ u^0_{\p} = \sqrt{1 + \left(u^3_{\p}\right)^2} 
                     \ .  
\end{eqnarray}
It thus follows that 
\beq
    \Lambda = \left \langle T^0_0 \right\rangle + \left \langle 
              T^z_z \right\rangle - \left\langle T^x_x \right\rangle 
              \ .
\eeq
We will use cylindrical coordinates and define $\phi_z = g_\omega 
\omega_z$ so that
\begin{eqnarray}
    m_* &=& m - {c_{\sigma}^2 \over 2 \pi^2}  m_* \left(
            \int_{- k_\n}^{k_\n} d k_z \left[k_\n^2 + \phi_z^2 + 
            m_*^2 + 2 \phi_z k_z\right]^{1/2} + 
            \int_{- k_\p}^{k_\p} d k_z \left[k_\p^2 + \left(
            \phi_z + K\right)^2 + m_*^2 + 2 \left(\phi_z + K\right) 
            k_z\right]^{1/2} - \right. \cr
         && \cr
         && \left.\int_{- k_\n}^{k_\n} d k_z \left[\left(k_z + \phi_z
            \right)^2 + m_*^2\right]^{1/2} - \int_{- k_\p}^{k_\p} 
            d k_z \left[\left(k_z + \phi_z + K\right)^2 
            + m_*^2\right]^{1/2}\right) \ , \\
         && \cr
    \n^{0} &=& {1 \over 3 \pi^2} k_{\n}^3 \quad , \quad
    \p^{0} = {1 \over 3 \pi^2} k_{\p}^3 \ , \\
         && \cr
    \n^{z} &=& {1 \over 2 \pi^2} \int_{- k_{\n}}^{k_{\n}} d k_{z} 
               \left(k_{z} + \phi_{z}\right) \left(\left[
               k_{\n}^2 + m_*^2 + \phi_{z}^2 + 2 \phi_{z} k_{z}
               \right]^{1/2} - \left[\left(k_{z} + \phi_{z}\right)^2 
               + m_*^2\right]^{1/2}\right) \ , \cr
         && \cr
    \p^{z} &=& {1 \over 2 \pi^2} \int_{- k_{\p}}^{k_{\p}} d k_{z} 
               \left(k_{z} + \phi_{z} + K\right) \left(\left[
               k_{\n}^2 + m_*^2 + \left(\phi_{z} + K\right)^2 + 
               2 \left(\phi_{z} + K\right) k_{z}\right]^{1/2} - 
               \right. \cr
            && \cr
            && \left.\left[\left(k_{z} + \phi_{z} + K\right)^2 
               + m_*^2\right]^{1/2}\right) \ . \label{cylcoords}
\end{eqnarray}
It is to be remarked that even though only the protons are given 
spatial momentum $K$ the neutron four-velocity nevertheless has a 
non-zero spatial component. \ This, in fact, is a signature of the 
entrainment effect, which is a momentum induced in one of the fluids 
will cause part of the other fluid to flow.

Of primary importance to the fluid equations are the $\A$, $\B$, and 
$\C$ coefficients. \ We could, in principle, use Eq.~(\ref{coordefs}) 
to express $(\n^2,\p^2,x^2)$ in terms of $(k_{\n},k_{\p},K)$, but 
practically speaking this is not possible. \ Fortunately we see from 
Eq.~(\ref{alg}) that we can construct these coefficients 
algebraically from the mean field values of the stress-energy tensor 
components. \ Thus, when it comes to the numerical work, we use the 
set $(k_{\n},k_{\p},K)$ as the independent variables. \ Note that 
because the master function is a scalar, it must be invariant if $K 
\to - K$, and is thus an even function of $K$.

\subsection{Equilibrium models}

The equilibrium configurations are spherically symmetric and static, 
so the metric can be written in the Schwarzschild form
\beq
  {\rm d}s^2 = - e^{\nu(r)} {\rm d}t^2 + e^{\lambda(r)} {\rm d}r^2 
               + r^2 \left({\rm d}\theta^2 + {\rm sin}^2\theta 
               {\rm d}\phi^2\right) \ . \label{bgmet}
\eeq
The two metric coefficients are determined from two Einstein 
equations, which are written as 
\beq
    \lambda^{\prime} = {1 - e^{\lambda} \over r} - 8 \pi r 
                         e^{\lambda} \left.\Lambda\right|_0 
                  \quad , \quad
    \nu^{\prime} = - {1 - e^{\lambda} \over r} + 8 \pi r e^{\lambda} 
                   \left.\Psi\right|_0 \ . \label{bckgrnd}
\eeq
The equations that determine the radial profiles of $\n(r)$ and 
$\p(r)$ have been derived by Comer et al.~\cite{CLL99} and they are  
\beq
    \left.\A^0_0\right|_0 \p^{\prime} + \left.\B^0_0\right|_0 
    \n^{\prime} + {1 \over 2} \left.\mu\right|_0 \nu^{\prime} = 0 
    \quad , \quad
    \left.\C^0_0\right|_0 \p^{\prime} + \left.\A^0_0\right|_0 
    \n^{\prime} + {1 \over 2} \left.\chi\right|_0 \nu^{\prime} = 0 
    \ , \label{bgndfl_c}
\eeq
where
\begin{eqnarray}
 \a00 &=& \A + 2 {\partial \B \over \partial \p^2} \n \p + 2 
          {\partial \A \over \partial \n^2} \n^2 + 2 {\partial \A 
          \over \partial \p^2} \p^2 + {\partial \A \over \partial 
          x^2} \p \n \ , \\
        && \cr
 \b00 &=& \B + 2 {\partial \B \over \partial \n^2} \n^2 + 4 
          {\partial \A \over \partial \n^2} \n \p + {\partial \A 
          \over \partial x^2} \p^2 , \\
        && \cr
 \c00 &=& \C + 2 {\partial \C \over\partial \p^2} \p^2 + 4 {\partial 
           \A \over \partial \p^2} \n \p + {\partial A \over \partial 
           x^2} \n^2 \ , \label{coef2}
\end{eqnarray}
and $\left.\Lambda\right|_0$, $\left.\Psi\right|_0$, 
$\left.\mu\right|_0$, and $\left.\chi\right|_0$ are given in the 
appendix. \ A zero subscript means that after the partial 
derivatives are taken, then one takes the limit $K \to 0$. 

Of course, since the variables that are more suited to the mean field 
theory are the two Fermi wavenumbers, $k_\n$ and $k_\p$, we replace 
everywhere $\n = k^3_\n/3 \pi^2$ and $\p = k^3_\p/3 \pi^2$ and solve 
for the wavenumbers instead. \ We have also found a more convenient 
way of determining the Dirac effective mass 
$\left.m_*\right|_0(k_\n,k_\p)$ and that is we have turned the 
transcendental algebraic relation in Eq.~(\ref{mstar}) of the 
appendix into a differential equation via
\beq
    \left.m^{\prime}_*\right|_0 = \left.{\partial m_* \over \partial 
     k_\n}\right|_0 k^{\prime}_\n + \left.{\partial m_* \over 
     \partial k_\p}\right|_0 k^{\prime}_\p \ ,
\eeq
where $k^{\prime}_\n$ and $k^{\prime}_\p$ are obtained from 
Eq.~(\ref{bgndfl_c}).   

The ``boundary'' conditions that must be imposed include a set at the 
center and another at the surface of the star. \ Demanding a 
non-singular behavior at the center of the star imposes that 
$\lambda(0) = 0$, and consequently that $\lambda^{\prime}(0)$ and 
$\nu^{\prime}(0)$ must also vanish. \ This and Eq.~(\ref{bgndfl_c}) 
imply that $k_\n^{\prime}(0)$ and $k_\p^{\prime}(0)$ have to vanish 
as well. \ A smooth joining of the interior spacetime to a 
Schwarzschild vacuum exterior at the surface of the star, 
i.e.~$r = R$, implies that the total mass $M$ of the system is 
given by
\beq
    M = - 4 \pi \int^{R}_0 {\rm d} r~r^2~\left.\Lambda\right|_0(r)
\eeq
and that $\left.\Psi\right|_0(R) = 0$.  

\subsection{The low velocity limit for fluid elements}

Two immediate applications of the formalism developed here are to 
model slowly rotating configurations \cite{AC01c} and linearized 
perturbations, or quasinormal modes \cite{CLL99,ACL02}. \ In both 
cases the fluid element velocities are small in the sense that they 
are typically only a few percent of the speed of light. \ The net 
effect is that these applications require only the first few terms 
from an expansion of the master function in terms of the entrainment 
parameter (i.e.~$x^2$ in the canonical formulation, and $K^2$ in what 
we present here). \ Such an expansion has been described in 
\cite{AC01a,ACL02}, and thus only the highlights will be reproduced 
here. \ It should be noted, however, that if one wanted to model 
rapidly rotating superfluid neutron stars \cite{PNC02}, say, then the 
expansion to be described below will be inappropriate. 

For a region within the fluid that is small enough that the 
gravitational field does not change appreciably across it one can 
show that 
\beq
    x^2 = \n \p \left({1 - \vec{v}_{\n} \cdot \vec{v}_{\p}/c^2 \over 
          \sqrt{1 - (v_{\n}/c)^2} \sqrt{1 - (v_{\p}/c)^2}}\right) \ .
\eeq
If it is the case that the individual three-velocities 
$\vec{v}_{\n ,\p}$ are small with respect to the speed of light, i.e.
\beq
     {v_{\n,\p} \over c} << 1 \ ,
\eeq
then it will be true that $x^2 \approx \n \p$ to leading order in the 
ratios $u_{\n}/c$ and $v_{\p}/c$. \ Thus, an appropriate expansion of 
the equation of state is 
\beq
    \Lambda(n^2,p^2,x^2) = \sum_{i = 0}^{\infty} \lambda_i(n^2,p^2) 
                           \left(x^2 - \n \p\right)^i \ ,
\eeq
since $x^2 - \n \p$ is small with respect to $\n \p$. \ In this case 
the $\A$, $\a00$, etc.~coefficients that appear in the field 
equations can be written as
\begin{eqnarray}
  \A &=& - \sum_{i = 1}^{\infty} i~\lambda_i(n^2,p^2) \left(x^2 - \n 
         \p\right)^{i - 1}  \ , \cr
     && \cr
  \B &=& - {1 \over \n} {\partial \lambda_0 \over \partial \n} - {\p 
         \over \n} \A - {1 \over \n} \sum_{i = 1}^{\infty} {\partial 
         \lambda_i \over \partial \n} \left(x^2 - \n \p\right)^i 
         \ , \cr
     && \cr
  \C &=& - {1 \over \p} {\partial \lambda_0 \over \partial \p} - {\n 
         \over \p} \A - {1 \over \p} \sum_{i = 1}^{\infty} {\partial 
         \lambda_i \over \partial \p} \left(x^2 - \n \p\right)^i 
         \ , \cr
     && \cr
  \a00 &=& - {\partial^2 \lambda_0 \over \partial \p \partial \n} - 
           \sum_{i = 1}^{\infty} {\partial^2 \lambda_i \over \partial 
           \p \partial \n} \left(x^2 - \n \p\right)^i \ , \cr
     && \cr
  \b00 &=& - {\partial^2 \lambda_0 \over \partial \n^2} - 
           \sum_{i = 1}^{\infty} {\partial^2 \lambda_i \over \partial 
           \n^2} \left(x^2 - \n \p\right)^i \ , \cr
     && \cr
  \c00 &=& - {\partial^2 \lambda_0 \over \partial \p^2} - 
           \sum_{i = 1}^{\infty} {\partial^2 \lambda_i \over \partial 
           \p^2} \left(x^2 - \n \p\right)^i \ .  
\end{eqnarray}

For quasinormal mode and slow-rotation calculations, each of the 
coefficients are evaluated on the background, so that $x^2 = \n \p$, 
and thus only the first few $\lambda_i$ are needed. \ In fact, one 
needs to retain only $\lambda_0$ and $\lambda_1$, where the latter 
contains the information concerning the entrainment effect. \ Some 
details are given in the appendix, and the final results are 
\begin{eqnarray}
\left.\A\right|_0 &=& c_{\omega}^2 + {c^2_{\omega} \over 5 
        \left.\mu^2\right|_0} \left(2 k^2_\p {\sqrt{k^2_\n + 
        \left.m^2_*\right|_0} \over \sqrt{k^2_\p + 
        \left.m^2_*\right|_0}} + {c^2_{\omega} \over 3 \pi^2} 
        \left[{k^2_\n k^3_\p \over \sqrt{k^2_\n + 
        \left.m^2_*\right|_0}} + {k^2_\p k^3_\n \over \sqrt{k^2_\p + 
        \left.m^2_*\right|_0} }\right]\right) + {3 \pi^2 k^2_{\p} 
        \over 5 \left.\mu^2\right|_0 k^3_\n} {k^2_\n + 
        \left.m^2_*\right|_0 \over \sqrt{k^2_\p + 
        \left.m^2_*\right|_0}} \ , \\
        && \cr
\left.\B\right|_0 &=& {3 \pi^2 \left.\mu\right|_0 \over k^3_\n} - 
        c_{\omega}^2 {k^3_\p \over k^3_\n} - {c^2_{\omega} k^3_\p 
        \over 5 \left.\mu^2\right|_0 k^3_\n} \left(2 k^2_\p 
        {\sqrt{k^2_\n + \left.m^2_*\right|_0} \over \sqrt{k^2_\p + 
        \left.m^2_*\right|_0}} + {c^2_{\omega} \over 3 \pi^2} 
        \left[{k^2_\n k^3_\p \over \sqrt{k^2_\n + 
        \left.m^2_*\right|_0}} + {k^2_\p k^3_\n \over \sqrt{k^2_\p + 
        \left.m^2_*\right|_0} }\right]\right) - \cr
        && \cr
        && {3 \pi^2 k^5_{\p} \over 5 \left.\mu^2\right|_0 k^6_\n} 
        {k^2_\n + \left.m^2_*\right|_0 \over \sqrt{k^2_\p + 
        \left.m^2_*\right|_0}} \ , \\
        && \cr
\left.\C\right|_0 &=& {3 \pi^2 \left.\chi\right|_0 \over k^3_\p} - 
        c_{\omega}^2 {k^3_\n \over k^3_\p} - {c^2_{\omega} k^3_\n 
        \over 5 \left.\mu^2\right|_0 k^3_\p} \left(2 k^2_\p 
        {\sqrt{k^2_\n + \left.m^2_*\right|_0} \over \sqrt{k^2_\p + 
        \left.m^2_*\right|_0}} + {c^2_{\omega} \over 3 \pi^2} 
        \left[{k^2_\n k^3_\p \over \sqrt{k^2_\n + 
        \left.m^2_*\right|_0}} + {k^2_\p k^3_\n \over \sqrt{k^2_\p + 
        \left.m^2_*\right|_0} }\right]\right) - \cr
        && \cr
        && {3 \pi^2 \over 5 \left.\mu^2\right|_0 k_\p} 
        {k^2_\n + \left.m^2_*\right|_0 \over \sqrt{k^2_\p + 
        \left.m^2_*\right|_0}} + {3 \pi^2 \over k^3_\p} 
        \sqrt{k^2_{\p} + m^2_e} \ , \\
        && \cr
\left.\a00\right|_0 &=& - {\pi^4 \over k^2_\n k^2_\p} \left.{
        \partial^2 \Lambda \over \partial k_\p \partial 
        k_\n}\right|_0  = c_\omega^2 + {\pi^2 \over k^2_\p} { 
        \left.m_*\right|_0 \left.{\partial m_* \over \partial k_\p}
        \right|_0 \over \sqrt{k^2_\n + \left.m^2_*\right|_0}}\ , \\
        && \cr
\left.\b00\right|_0 &=& {\pi^4 \over k^5_\n} \left(2 
        \left.{\partial \Lambda \over \partial k_\n}\right|_0 - 
        k_\n \left.{\partial^2 \Lambda \over \partial k^2_\n}
        \right|_0\right) = c_\omega^2 + {\pi^2 \over k^2_\n} {k_\n + 
        \left.m_*\right|_0 \left.{\partial m_* \over \partial k_\n}
        \right|_0 \over \sqrt{k^2_\n + \left.m^2_*\right|_0}} \ , \\
        && \cr
\left.\c00\right|_0 &=& {\pi^4 \over k^5_\p} \left(2 \left.{\partial 
        \Lambda \over \partial k_\p}\right|_0 - k_\p 
        \left.{\partial^2 \Lambda \over \partial k^2_\p}
        \right|_0\right) = c_\omega^2 + {\pi^2 \over k^2_\p} {k_\p + 
        \left.m_*\right|_0 \left.{\partial m_* \over \partial k_\p}
        \right|_0 \over \sqrt{k^2_\p + \left.m^2_*\right|_0}} + 
        {\pi^2 \over k_\p} {1 \over \sqrt{k^2_\p + m^2_e}} \ , 
\end{eqnarray}
where $\left.\partial m_*/\partial k_\n\right|_0$ and $\left.\partial 
m_*/\partial k_\p\right|_0$ can be found in the appendix and $m_e = 
m/1836$. \ For reasons to be discussed below, we have included 
contributions due to a normal fluid of highly degenerate electrons. 

\subsection{Equilibrium configurations}

\begin{figure}[t]
\centering
\vskip 24pt
\includegraphics[height=6.5cm,clip]{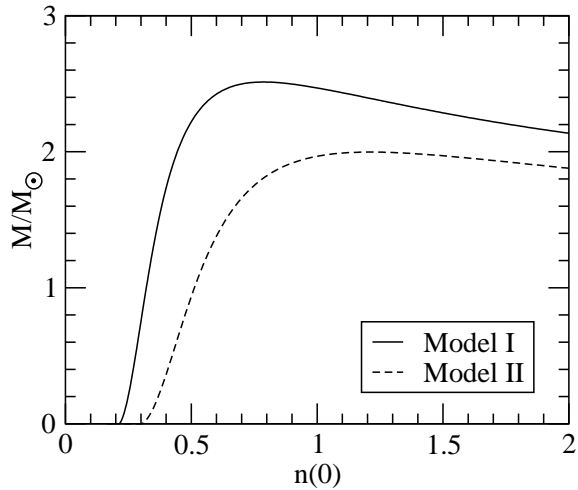}
\caption{Mass $M$ (in units of solar mass $M_{\odot}$) vs the central 
neutron number density $n(0)$ (in units of ${\rm fm}^{- 3}$) for the 
mean field coupling values of Models I and II of Table~\ref{table1}.}
\label{mass1}
\end{figure}

We now use our model to construct static and spherically symmetric 
configurations. \ A priori there are two input parameters, which are 
the neutron and proton Fermi wavenumbers at the center of the star.  
However, we can reduce this to just the neutron wavenumber by 
imposing the condition of chemical equilibrium between the nucleons. 
\ In order to have a chemical equilibrium that is believed to be 
representative of neutron stars (i.e.~proton fractions $x_\p = 
\p/(\n + \p) \approx 0.1$), we have added to the master function a 
term (see, for instance, \cite{ST83,ACP02b}) that accounts for a 
highly degenerate gas of relativistic leptons (in our case, just 
electrons). \ Fig.~\ref{mass1} gives the mass $M$ as a function of 
the central neutron number density $\n(0)$. \ We see the typical 
behaviour of general relativistic neutron stars, and that is a 
maximum value for the mass. \ Beyond this maximum, the stars will be 
in unstable equilibria. \ As canonical models of superfluid neutron 
stars, we have chosen configurations that are near to the maximum 
mass, but on the stable branch of the curves (cf.~Table 
\ref{table1}).    

\begin{table}
\caption{Parameters describing our choice of mean field and canonical 
superfluid neutron star models. \ The two values for $c^2_{\sigma}$ 
and $c^2_{\omega}$ represent the two extremes given in \cite{G97} 
that have been determined from nuclear physics. \ Note that the 
baryon mass is $m = 4.7582~{\rm fm}^{- 1}$.}
\begin{tabular}{|c|c|}
\hline
  Model I & Model II \cr
\hline
  $c_\sigma^2 = 12.684$ & $c_\sigma^2 =8.403$ \cr
  $c_\omega^2 = 7.148$ & $c_\omega^2 = 4.233$ \cr  
  $\nu(0) = - 2.316408$ & $\nu(0) = - 2.288385$ \cr 
  $k_\n(0) = 2.8~{\rm fm}^{- 1}$ & $k_\n(0) = 3.25~{\rm fm}^{- 1}$ \cr
  $x_\p(0) = 0.101$ & $x_\p(0) = 0.102$ \cr
  $M = 2.509~M_\odot$ & $M = 1.996~M_\odot$ \cr 
  $R = 11.696~{\rm km}$ & $R = 9.432~{\rm km}$ \cr
\hline
\end{tabular}
\label{table1}
\end{table}

\begin{figure}[t]
\centering
\vskip 24pt
\includegraphics[height=6.5cm,clip]{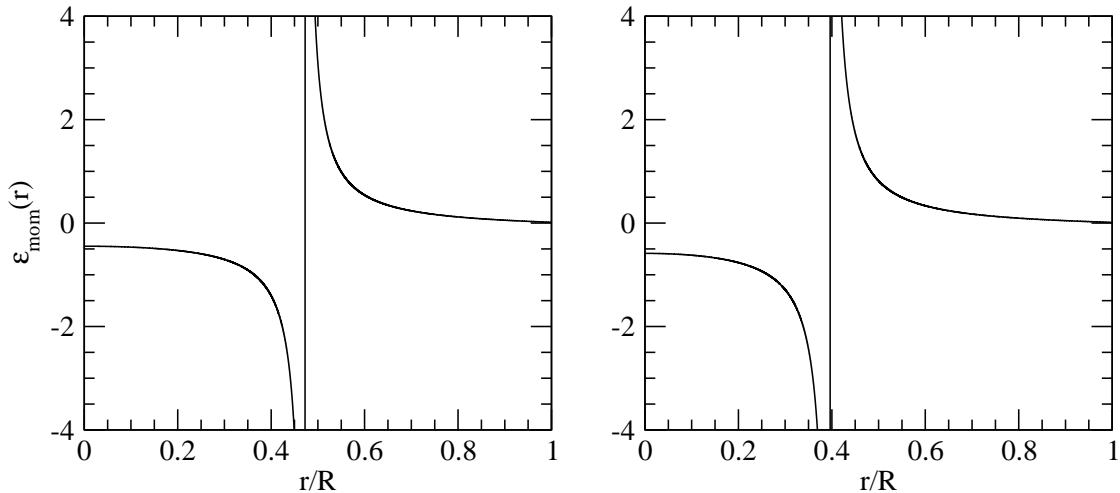}
\caption{The entrainment parameter $\varepsilon_{mom}$ as a function 
of radius for Models I (left panel) and II (right panel) of 
Table~\ref{table1}.}
\label{eps1}
\end{figure}

In several earlier studies \cite{LM00,PCA02,ACL02}, a parameterized 
model for entrainment has been used that is based on the Newtonian 
calculations of Borumand et al.~\cite{BJK96}. \ The parameter, 
$\varepsilon_{mom}$, in this model can be shown to be related to our 
coefficient $\left.\A\right|_0$ via 
\beq
    \varepsilon_{mom} = {\left.\A\right|_0 \p \over m - 
    \left.\A\right|_0 \left(\n + \p\right)} \ . 
\eeq
In the previous studies the ``physical'' range has been taken to be 
$0.04 < \varepsilon_{mom} < 0.2$. \ But already with this formula we 
find that $\varepsilon_{mom}$ has a singularity whenever $m = \left.
\A\right|_0 (\n + \p)$. \ And in Fig.~\ref{eps1} we see that 
indeed $\varepsilon_{mom}$ has a discontinuity in both curves (which 
are for Models I and II of Table~\ref{table1}). \ This is perhaps the 
most significant short term result of this work, and hence taking 
$\varepsilon_{mom}$ to be a constant must be considered a non-viable 
option.    

\begin{figure}[t]
\centering
\vskip 24pt
\includegraphics[height=6.5cm,clip]{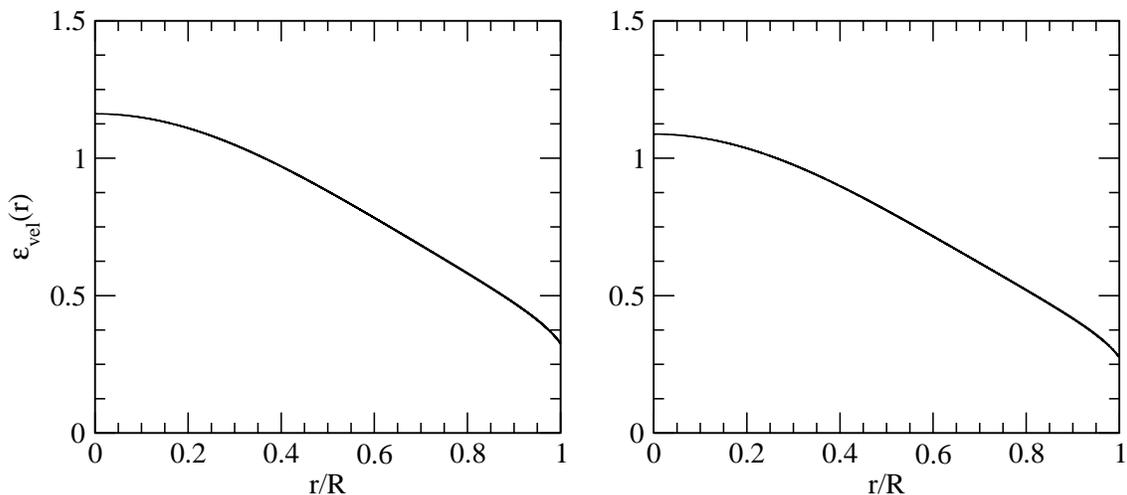}
\caption{The entrainment parameter $\varepsilon_{vel}$ as a function 
of radius for Models I (left panel) and II (right panel) of 
Table~\ref{table1}.}
\label{eps2}
\end{figure}

An alternative parameterization is that of Prix et al.~\cite{PCA02}.  
They point out that there is some ambiguity in what is meant by the 
nucleon effective masses (i.e.~the Landau, as opposed to the Dirac, 
effective masses \cite{G97}), which can be traced to whether one 
chooses to define these masses with respect to the zero-momentum or 
zero-velocity frame of the nucleons. \ Their parameterization makes 
use of the zero-velocity frame. \ In this case one finds   
\beq
    \varepsilon_{vel} = {\left.\A\right|_0 \n \over m} \ . 
\eeq
Immediately we see that this formulation does not have an obvious 
singularity anywhere, and this is reflected in Fig.~\ref{eps2} which 
contains plots of the radial profile of $\varepsilon_{vel}$ for 
Models I and II of Table~\ref{table1}. \ The ``physical'' range has 
been taken to be $0.4 < \varepsilon_{vel} < 0.7$. \ We see from the 
figure that this is the range in the outer portion of the star, but 
in the core the entrainment stretches well out of this range.

Also used in Prix et al.~\cite{PCA02} is the so-called symmetry 
energy parameter 
\beq
    \sigma = {\partial \mu/\partial \p \over \partial \chi/\partial 
             \n} 
\eeq
that can be related to terms \cite{PAL88} in the equation of state 
that tend to force an equal number of protons and neutrons (as in 
most nuclei). \ For the relativistic mean field model used here, it 
is not difficult to show that $\sigma = 1$ (which is consistent with 
the range of values used by \cite{PCA02}). 

\section{Slow Rotation Configurations: The Frame-Dragging} 
\label{frame}

The key distinguishing feature of slowly rotating superfluid neutron 
stars is that the neutrons can rotate at rates different from that of 
the protons. \ The slow-rotation approximation is valid when the 
angular velocities are small enough that the fractional changes in 
pressure, energy density, and gravitational field due to the rotation 
are all relatively small. \ This translates into the inequalities 
(cf.~\cite{H67,AC01c})
\beq
    \Omega^2_{\n}~{\rm or}~\Omega^2_{\p}~{\rm or}~\Omega_{\n} 
    \Omega_{\p}~<<~\left({c \over R}\right)^2 {G M \over R c^2} \ ,
\eeq  
where the speed of light $c$ and Newton's constant $G$ have been 
restored, and $R$ and $M$ are the radius and mass, respectively, of 
the non-rotating configuration. \ Since $G M/c^2 R < 1$, we also see 
that  
\beq
    \Omega_{\n} R << c \qquad {\rm and} \qquad \Omega_{\p} R << c \ , 
\eeq
and thus the slow-rotation approximation ought to be useful for most 
astrophysical neutron stars. \ In fact a comparison 
\cite{AC01c,PCA02} of the above conditions to empirical estimates for 
the Kepler frequency (i.e.~the rotation rate at which mass-shedding 
sets in at the equator) that can been obtained from calculations 
using realistic supranuclear equations of state reveals that even the 
fastest observed pulsars can be classified as slowly rotating.

The only quantities that contain terms linear in the angular 
velocities are the metric coefficient $\omega(r)$, that represents 
the dragging of inertial frames, and the fluid four-velocities. \ All 
other effects due to rotation enter at the second-order in the 
angular velocities. \ It is useful to define 
\beq
    \tilde{L}_{\n} = \omega - \Omega_{\n} \quad , \quad 
    \tilde{L}_{\p} = \omega - \Omega_{\p} \ .
\eeq
Up to an overall minus sign, these represent rotation frequencies as 
perceived by local zero-angular momentum observers. \ The Einstein 
equation that determines the frame-dragging has been shown to be 
\cite{AC01c}
\beq
    {1 \over r^4} \left(r^4 e^{- (\lambda + \nu)/2} 
    \tilde{L}_{\p}^{\prime}\right)^{\prime} - 16 \pi e^{(\lambda - 
    \nu)/2} \left(\left.\Psi\right|_0 - \left.\Lambda\right|_0\right) 
    \tilde{L}_{\p} = 16 \pi e^{(\lambda - \nu)/2} \left.\mu\right|_0 
    \n \left(\Omega_{\n} - \Omega_{\p}\right) \ . \label{frmdrg}
\eeq
It is of the same form as that obtained by Hartle \cite{H67} except 
for the non-zero source term on the right-hand-side.  

Exterior to the star, there is vacuum, and so the solution for the 
frame-dragging is the same as that considered by Hartle \cite{H67}, 
i.e.
\beq
   \omega(r) = {2 J \over r^3} \ . \label{fvac}
\eeq
Assuming that the frame-dragging is continuous at the surface of the 
star,  then  
\beq
    J = - {8 \pi \over 3} \int_0^R {\rm d}r r^4 e^{(\lambda - \nu)/2}
        \left[\left.\mu\right|_0 \n \tilde{L}_{\n} + 
        \left.\chi\right|_0 \p \tilde{L}_{\p}\right] \ , 
\eeq
where $J$ is the total angular momentum. \ Andersson and Comer 
\cite{AC01c} have furthermore shown that the neutron total angular 
momentum is 
\beq
J_{\n} = - {8 \pi \over 3} \int_0^R {\rm d}r r^4 e^{(\lambda - \nu)/2}
         \left[\left.\mu\right|_0 \n \tilde{L}_{\n} + 
         \left.\A\right|_0 \n \p \left(\Omega_{\n} - \Omega_{\p}
         \right)\right]
\eeq
and 
\beq
J_{\p} = - {8 \pi \over 3} \int_0^R {\rm d}r r^4 e^{(\lambda - \nu)/2}
         \left[\left.\chi\right|_0 \p \tilde{L}_{\p} + 
         \left.\A\right|_0 \n \p \left(\Omega_{\p} - \Omega_{\n}
         \right)\right]
\eeq
for the proton total angular momentum, from which it follows that  
\beq
    J = J_{\n} + J_{\p} \ .
\eeq

Any solution of Eq.~(\ref{frmdrg}) for the frame-dragging is to be 
such that the interior matches smoothly onto the known vacuum 
solution in Eq.~(\ref{fvac}). \ This means that we must have, for 
instance, 
\beq
    \tilde{L}_{\p}(R) = - \Omega_\p + {2 J \over R^3} \ . 
    \label{surfc1}
\eeq
We can easily see that $\tilde{L}_\p$ and its derivative are 
smooth provided that we have
\beq
    \tilde{L}_{\p}(R) = - \Omega_\p - {R \over 3} \left.{d 
                        \tilde{L}_{\p} \over d r}\right|_{r = R} 
                        \ . \label{ltsurf}
\eeq 
Having obtained a value for $\tilde{L}_\p(0)$ that satisfies 
Eq.~(\ref{ltsurf}), an acceptable solution is in hand, and we can 
thus  determine the angular momentum of the configuration from 
Eq.~(\ref{surfc1}). 

\begin{figure}[t]
\centering
\vskip 24pt
\includegraphics[height=15cm,clip]{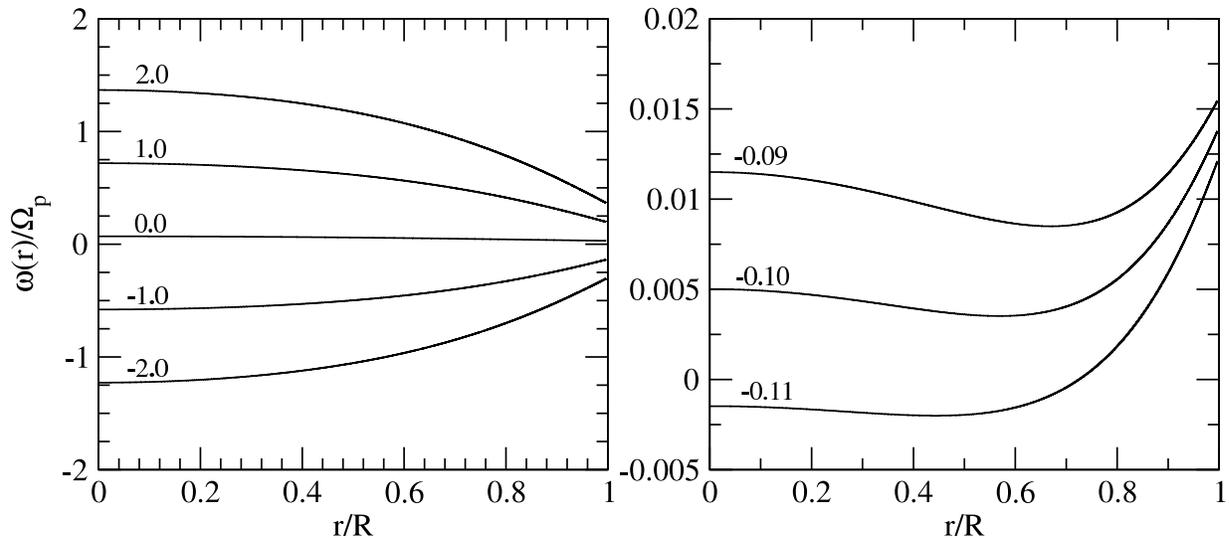}
\caption{The radial profile of the frame-dragging $\omega(r)$ for 
Model I of Table~\ref{table1}. \ In the left-panel we have curves for 
$\Omega_\n/\Omega_\p = (-2.0,-1.0,0.0,1.0,2.0)$, and in the right we 
have taken $\Omega_\n/\Omega_\p = (-0.11,-0.10,-0.09)$.}
\label{frames3}
\end{figure}

In Fig.~\ref{frames3} we have plots of the radial profile of the 
frame-dragging for Model I for a range of values of the ratio 
$\Omega_\n/\Omega_\p$. \ For the values considered in the left panel 
we see that the frame-dragging is much like that of an ordinary 
one-fluid star, and is consistent with solutions obtained by 
Andersson and Comer \cite{AC01c}. \ For the negative ratios, we see 
that the frame-dragging is negative but increases monotonically 
towards zero. \ This is the behaviour we should expect, since the 
bulk of the matter is simply rotating the opposite way. \ There is 
some asymmetry between the negative and positive ratios, but that is 
due to the small number of protons that rotate oppositely to the 
neutrons when the ratio is negative. \ In the right panel, we examine 
the solutions near to a ratio of zero. \ The frame-dragging is no 
longer monotonic and actually becomes negative inside the star. \ An 
explanation of this can be understood as follows: in the interior the 
protons carry most of the angular momentum and thus have the largest 
impact on the frame-dragging, but further away from the center, the 
much larger mass contained in the neutrons begins to dominate 
\cite{AC01c}.

\begin{figure}[t]
\centering
\vskip 24pt
\includegraphics[height=6.5cm,clip]{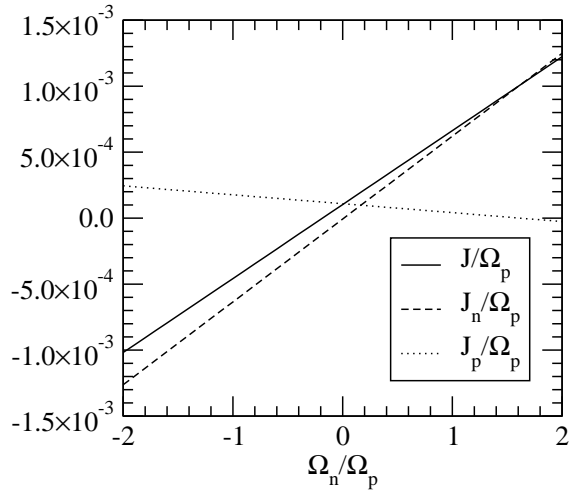}
\caption{The neutron $J_\n$, proton $J_\p$, and total $J$ angular 
momenta vs the ratio $\Omega_\n/\Omega_\p$ for Model I of 
Table~\ref{table1}.}
\label{angmom}
\end{figure}

Fig.~\ref{angmom} considers the same range for the ratio of the 
angular speeds, by showing how the total angular momentum $J$, and 
the neutron and proton angular momenta, $J_\n$ and $J_\p$ 
respectively, vary as $\Omega_\n/\Omega_\p$ is changed. \ As one 
might expect, when the ratio becomes greater than one, then the 
angular momentum in the neutrons is significantly greater than that 
of the protons. \ Likewise, as the ratio becomes smaller we find that 
the protons can dominate. \ Something of a surprise is the extreme 
right-hand-side of the curves, where $J_\p$ actually becomes 
negative, and yet the angular speeds of the neutrons and protons have 
the same sign. \ We can explain this behaviour as a purely general 
relativistic effect that is intimately connected with the 
frame-dragging. \ With respect to infinity, particles that are 
rotating around at the same rate as the local inertial frames are 
found to have zero angular momentum. \ Thus, those particles that 
would be lagging behind the frames, even though their angular 
trajectories would be in the same direction as the frames, will 
nevertheless have negative angular momentum. \ Finally, one other  
feature is the configuration where $J_\n = J_\p$. \ In this case, the 
angular speeds are not equal, nor are the total neutron and proton 
particle numbers equal, and yet the angular momenta of both fluids 
are the same.  

\section{Conclusions}

We have developed a formalism that uses relativistic mean field 
theory for supra-nuclear density matter that can be applied to 
general relativistic superfluid neutron stars. \ In this formalism 
we have also allowed for the entrainment effect between the various 
superfluid species. \ We have also shown how to use our formalism in 
the relativistic superfluid field equations that have been recently 
developed for modelling slowly rotating equilbirium configurations 
\cite{AC01c}, as well as the linearized oscillations 
\cite{CLL99,ACL02}. \ Our results indicate that parameterized models 
of entrainment that use a zero-momentum rest-frame definition for 
nucleon effective masses have a singularity, whereas those based on 
zero-velocity rest-frames do not.  

Our results should find a wide range of applications, not the least 
of which is to understand better the role of entrainment in the 
superfluid modes of oscillation (e.g.~the avoided crossings described 
by \cite{AC01a}) and subsequent imprints \cite{AC01b,GC02} that may 
be left in neutron star gravitational waves (emitted, for instance, 
during glitches). \ Applications planned for the near future will 
include numerical studies of rapidly rotating superfluid neutron 
stars (using an adaptation of the very accurate LORENE code 
\cite{PNC02}) and continued research on the newly discovered 
two-stream instability \cite{ACP02a,ACP02b}, which has been proposed 
as a trigger mechanism for glitches in pulsars. \ For the rapid 
rotation calculations one must necessarily imploy the full formalism 
discussed here in the sense that $K$ will no longer be kept small, 
since the LORENE code is specifically designed to accurately handle 
relative velocities of the neutrons with respect to the protons that 
approach the speed of light. \ And for the two-stream instability 
entrainment provides one of the main couplings between the two fluids.

\acknowledgements

We like to thank N.~Andersson and R.~Prix for useful input at various 
stages of this work.  GLC gratefully acknowledges partial support 
from NSF Gravitational Theory, Grant No.~PHYS-0140138, a 
Saint Louis University SLU2000 Faculty Research Leave award, and 
EPSRC grant GR/R52169/01 in the UK. \ RJ gratefully acknowledges 
partial support from the NSF Materials Theory Program, Grant 
No.~DMR-0081039.

\section*{Appendix: Limiting Forms}

The slow-rotation approximation is such that only terms up to and 
including ${\cal O}(\Omega^2_{\n,\p})$ are required. \ This 
translates into keeping only those terms in the mean field theory up 
to and including ${\cal O}(K^2)$. \ This is because those quantities 
like $m_*$ and $\Lambda$ are scalars, and can only depend on terms 
that are even in $K$. \ Likewise, those quantities that are like 
vectors, e.g.~$\omega_z$, can only depend on terms that are odd in 
$K$. \ Because $m_*$ and $\omega_z$ are known only implicitly, we 
determine their expansion coefficients by assuming they take the form
\begin{eqnarray}
    \phi_z &=& \left.{\partial \phi_z \over \partial K}\right|_0 K 
               \ , \cr 
          && \cr
    m_* &=& \left.m_*\right|_0 + \left.{\partial m_* \over \partial 
            K^2}\right|_0 K^2 \ , \label{expans}
\end{eqnarray}
where 
\begin{eqnarray}
    \left.m_*\right|_0 &=& m_*(k_\n,k_\p,0) \cr
         && \cr
        &=& m - \left.m_*\right|_0 {c_\sigma^2 \over 2 \pi^2}  
            \left(k_\n \sqrt{k^2_\n + \left.m^2_*\right|_0} + k_\p 
            \sqrt{k^2_\p + \left.m^2_*\right|_0} + {1 \over 2} 
            \left.m_*^2\right|_0 {\rm ln} \left[{- k_\n + 
            \sqrt{k^2_\n + \left.m^2_*\right|_0} \over k_\n + 
            \sqrt{k^2_\n + \left.m^2_*\right|_0}}\right] + \right. \cr
         && \cr
         && {1 \over 2} \left.\left.m_*^2\right|_0 {\rm ln} \left[{- 
            k_\p + \sqrt{k^2_\p + \left.m^2_*\right|_0} \over k_\p + 
            \sqrt{k^2_\p + \left.m^2_*\right|_0}}\right]\right) \ . 
            \label{mstar}
\end{eqnarray}
By inserting Eq.~(\ref{expans}) into Eq.~(\ref{cylcoords}), and 
expanding and keeping terms to the appropriate orders, we find 
\begin{eqnarray}
     \left.{\partial \phi_z \over \partial K}\right|_0 &=& - 
          {c_\omega^2 \over 3 \pi^2} {k^3_\p \over \sqrt{k^2_\p + 
          \left.m^2_*\right|_0}} \left(1 + {c_\omega^2 \over 3 \pi^2} 
          \left[{k^3_\n \over \sqrt{k^2_\n + \left.m^2_*\right|_0}} + 
          {k^3_\p \over \sqrt{k^2_\p + \left.m^2_*\right|_0}}\right]
          \right)^{- 1} \ , \\
          && \cr
     \left.{\partial m_* \over \partial k_\n}\right|_0 &=& - 
          {c_\sigma^2 \over \pi^2} {\left.m_*\right|_0 k^2_\n 
          \over \sqrt{k^2_\n + \left.m^2_*\right|_0}} \left({3 m - 2 
          \left.m_*\right|_0 \over\left.m_*\right|_0} - {c_\sigma^2 
          \over \pi^2} \left[{k^3_\n \over \sqrt{k^2_\n + \left.m^2_*
          \right|_0}} + {k^3_\p \over \sqrt{k^2_\p + 
          \left.m^2_*\right|_0}}\right]\right)^{- 1} 
          \ , \\
          && \cr
     \left.{\partial m_* \over \partial k_\p}\right|_0 &=& - 
          {c_\sigma^2 \over \pi^2} {\left.m_*\right|_0 k^2_\p \over 
          \sqrt{k^2_\p + \left.m^2_*\right|_0}} \left({3 m - 2 
          \left.m_*\right|_0 \over\left.m_*\right|_0} - {c_\sigma^2 
          \over \pi^2} \left[{k^3_\n \over \sqrt{k^2_\n + 
          \left.m^2_*\right|_0}} + {k^3_\p \over \sqrt{k^2_\p + 
          \left.m^2_*\right|_0}}\right]\right)^{- 1} 
          \ , \\
          && \cr    
      \n^{z} &=& {1 \over 3 \pi^2} {k^3_\n \over \sqrt{k^2_\n + 
          \left.m^2_*\right|_0}} \left.{\partial \phi_z \over 
          \partial K}\right|_0 K \ , \\
          && \cr
      \p^{z} &=& {1 \over 3 \pi^2} {k^3_\p \over \sqrt{k^2_\p + 
             \left.m^2_*\right|_0}} \left(\left.{\partial \phi_z 
             \over \partial K}\right|_0 + 1\right) K \ . 
\end{eqnarray}
We note that the coefficient $\left.\partial m_*/\partial K^2
\right|_0$ cancels everywhere, which is why it is not written here. \ 
Also, we find 
\begin{eqnarray}
    \left.\Lambda\right|_0 &=& - {1 \over 4 \pi^2} \left(k_\n^3 
             \sqrt{k^2_\n + \left.m^2_*\right|_0} + 
             k_\p^3 \sqrt{k^2_\p + \left.m^2_*\right|_0}
             \right) - {1 \over 2} m^2_\omega \omega_0^2 - {1 \over 
             4} c_\sigma^{- 2} \left(2 m - \left.m_*\right|_0\right) 
             \left(m - \left.m_*\right|_0\right) - \cr
             && \cr
             &&{1 \over 8 \pi^2} \left(m_e k_\p \left[2 k_\p + m_e
             \right] \sqrt{k_\p^2 + m^2_e} - m^4_e {\rm ln}\left[
             {k_\p + \sqrt{k^2_\p + m^2_e} \over m_e}\right]\right) 
             \ , \\
             && \cr
    \left.\mu\right|_0 &=& g_\omega \omega_0 + \sqrt{k^2_\n + 
             \left.m^2_*\right|_0} \ , \\ 
             && \cr
    \left.\chi\right|_0 &=& g_\omega \omega_0 + \sqrt{k^2_\p + 
             \left.m^2_*\right|_0} \  , \\ 
             && \cr
    \left.\Psi\right|_0 &=& \left.\Lambda\right|_0 + {1 \over 3 \pi^2}
             \left(\left.\mu\right|_0 k_{\n}^3 + \left[
             \left.\chi\right|_0 + \sqrt{k_{\p}^2 + 
             m^2_e}\right]k_{\p}^3\right) \ .
\end{eqnarray}
The condition of chemical equilibrium is that $\left.\mu\right|_0 = 
\left.\chi\right|_0 + \sqrt{k^2_\p + m^2_e}$.

\end{document}